\documentclass[aps,prl,twocolumn,superscriptaddress,showpacs]{revtex4}
\usepackage{epsfig}
\usepackage[dvipsnames,usenames]{color}
\usepackage{color}
\usepackage{amsmath}

\emergencystretch=\maxdimen
\begin{document}
\title{Universal scaling in the Knight shift anomaly of doped periodic Anderson model}
\author{M. Jiang}
\email[]{jiangmi@itp.phys.ethz.ch}
\affiliation{Institute for Theoretical Physics, ETH Zurich, Switzerland}
\author{Yi-feng Yang}
\email[]{yifeng@iphy.ac.cn}
\affiliation{Beijing National Laboratory for Condensed Matter Physics and Institute of Physics, Chinese Academy of Sciences, Beijing, China}
\address{Collaborative Innovation Center of Quantum Matter, Beijing 100190, China}
\address{School of Physical Sciences, University of Chinese Academy of Sciences, Beijing 100190, China}

\begin{abstract}
We report a Dynamical Cluster Approximation (DCA) investigation of the doped periodic Anderson model (PAM) to explain the universal scaling in the Knight shift anomaly predicted by the phenomenological two-fluid model and confirmed in many heavy-fermion compounds. We calculate the quantitative evolution of the orbital-dependent magnetic susceptibility and reproduce correctly the two-fluid prediction in a large range of doping and hybridization. Our results confirm the presence of a temperature/energy scale $T^{\ast}$ for the universal scaling and show distinctive behavors of the Knight shift anomaly in response to other ``orders'' at low temperatures. However, comparison with the temperature evolution of the calculated resistivity and quasiparticle spectral peak indicates a different characteristic temperature from $T^*$, in contradiction with the experimental observation in CeCoIn$_5$ and other compounds. This reveals a missing piece in the current model calculations in explaining the two-fluid phenomenology.
\end{abstract}

\maketitle

Heavy-fermion materials, which typically contain Ce, Yb, U, or Pu ions, exhibit complex behaviors arising from interplay between localized and itinerant electrons \cite{zachreview,stewartreviewRMP1984,Yangreview}. These include exotic ground states such as the Q-phase with coexisting magnetism and superconductivity in CeCoIn$_5$, the so-called ``hidden'' order in URu$_2$Si$_2$, and unconventional superconductivity in these and many other materials \cite{Thompson2001,MydoshURSreview2011}. A common basis of these lies in the breakdown of conventional Fermi-liquid theory in proximity to a quantum phase transition
\cite{YRSnature,ColemanQCreview,StewartHFreview}. The recent discovery of the CeMIn$_5$ (M = Rh, Ir, or Co) class of materials has continually highlighted the requirement of a general understanding of the heavy-electron phenomena, especially their universal behaviors \cite{tuson,Young2007,KenzelmannCeCoIn5Qphase}. In particular, the universal nuclear magnetic resonance (NMR) Knight shift anomaly has attracted considerable attention \cite{Curroreview} among various exotic behaviors.

NMR probes the relative shift of nuclear resonance frequency compared with that of isolated nucleus due to hyperfine coupling between nuclear and electron spins. The Knight shift of a normal Fermi liquid is given by $K = A\chi_{0}$, where $A$ is the hyperfine coupling and $\chi_{0}$ is the Pauli susceptibility proportional to the density of states at the Fermi energy, so that $K\propto AN(0)$ is approximately temperature independent. The proportionality between the Knight shift and the susceptibility also appears for a pure spin system, where $\chi_0$ measures the Curie-Weiss susceptibility of the localized spins. However, this fails to describe the unusual normal state property of heavy-fermion materials. While the Knight shift $K$ is indeed proportional to the susceptibility at high temperature, the simple relation fails below a material-dependent crossover temperature $T^{\ast}\sim 10-100$ K, which reflects the onset of lattice coherence or hybridization between conduction electrons and localized $f$-electrons. This Knight shift anomaly has been detected in dozens of heavy-fermion materials, including the CeMIn$_5$ family, CeCu$_2$Si$_2$, UPt$_3$, and URu$_2$Si$_2$ \cite{Curro04,KentCurro}. Previous attempts to explain the Knight shift anomaly either argue for a temperature dependent hyperfine interaction due to Kondo screening \cite{theory1}, or attribute to the temperature variation of the crystal field occupations of the 4f(5f) electrons in the rare earth or actinide ions \cite{theory2}. However, there is evidence that the hyperfine coupling has normally a much higher energy scale than the Kondo and/or crystal field interactions so that their role in giving rise to the dramatic changes observed experimentally is in question~\cite{theory3}.

Recently, a phenomenological two-fluid theory has emerged as a promising framework in explaining diverse non-Fermi liquid behaviors observed in heavy-electron materials~\cite{Yangreview,twofluid2004,twofluid2008,twofluid2008a,twofluid2012,twofluid2016}.
This two-fluid model argues that below a coherence temperature, $T^*$, an itinerant heavy-electron Kondo liquid, which displays non-Fermi liquid scaling behavior, emerges as a new quantum state of matter as the localized $f$-electrons collectively reduce their entropy through collective hybridization with the conduction electrons. This emergent Kondo liquid coexists with a spin liquid formed by the lattice of local moments of $f$-electrons whose magnitude is reduced by the hybridization. At low temperatures, both the residual unhybridized local moments and the emergent Kondo liquid interplay to determine the ground state of the system such as antiferromagnetism, unconventional superconductivity, or other competing orders~\cite{ColemanUnderscreenedPRL1992,twofluid2004,twofluid2008,twofluid2008a,twofluid2012,twofluid2016}.  Although the  intertwined two-fluid picture has gained much interest in successfully accounting for many anomalous heavy-fermion properties, a minimal microscopic model that can clarify the nature of the two fluids and provide a comprehensive and quantitative understanding of the phenomenological theory is still lacking~\cite{twofluid2016}.

The periodic Anderson model (PAM), which consists of a lattice of strongly correlated $f$-electrons embedded in a background of conduction electrons, is believed to capture the essence of heavy-fermion physics. The PAM Hamiltonian reads:
\begin{eqnarray}
    {\cal H} = &-&t \sum\limits_{\langle ij \rangle,\sigma}
(c^{\dagger}_{i\sigma}c_{j\sigma}^{\vphantom{dagger}}
+c^{\dagger}_{j\sigma}c_{i\sigma}^{\vphantom{dagger}})
        -V \sum\limits_{i\sigma}
(c^{\dagger}_{i\sigma}f_{i\sigma}^{\vphantom{dagger}}+
f^{\dagger}_{i\sigma}c_{i\sigma}^{\vphantom{dagger}})
\nonumber \\
        &+& U \sum\limits_{i} (n^{f}_{i\uparrow}-\frac{1}{2})
(n^{f}_{i\downarrow}-\frac{1}{2})
\label{eq:PAM}
\end{eqnarray}
where $c^{\dagger}_{i\sigma}(c_{i\sigma}^{\vphantom{dagger}})$ and $f^{\dagger}_{i\sigma}(f_{i\sigma}^{\vphantom{dagger}})$
are creation(annihilation) operators for the conduction and local $f$-electrons on site $i$ with spin $\sigma$, respectively. $n^{c,f}_{i\sigma}$ are the associated number operators. Here for simplicity the chemical potential for the conduction electron site energy is fixed to $\mu=E_f=0$, while the $f$-electron band is doped for a desired occupancy covering both the Kondo regime ($\langle n^f_i\rangle\approx 1$) and the mixed-valence regime. $t$ is the hopping parameter of the conduction electrons on nearest-neighbor sites $\langle ij \rangle$ of a square lattice, $U$ the local repulsive interaction in the $f$-orbital and $V$ the hybridization between the conduction and $f$-electrons. $t=1$ sets the energy scale throughout the paper. At low temperatures, the hybridization $V$ between conduction and $f$-electrons and repulsive interaction $U$ for $f$-electrons compete to determine the ground state, which reflects the competition between the Ruderman-Kittel-Kasuya-Yosida (RKKY) scale, $\sim J^{2}/W$, and the Kondo energy scale, $\sim W e^{-W/J}$, with $J\sim V^{2}/U$ and $W$ the bandwidth of the conduction electrons~\cite{PAMRTS,PAMreference}. Specifically, at small $V$, local $f$-moments couple antiferromagnetically via the indirect RKKY interaction mediated by the conduction band; while at large $V$, the conduction and $f$-electrons bind into a global spin singlet via the Kondo interaction, giving rise to a paramagnetic ground state.

\begin{figure}
\psfig{figure=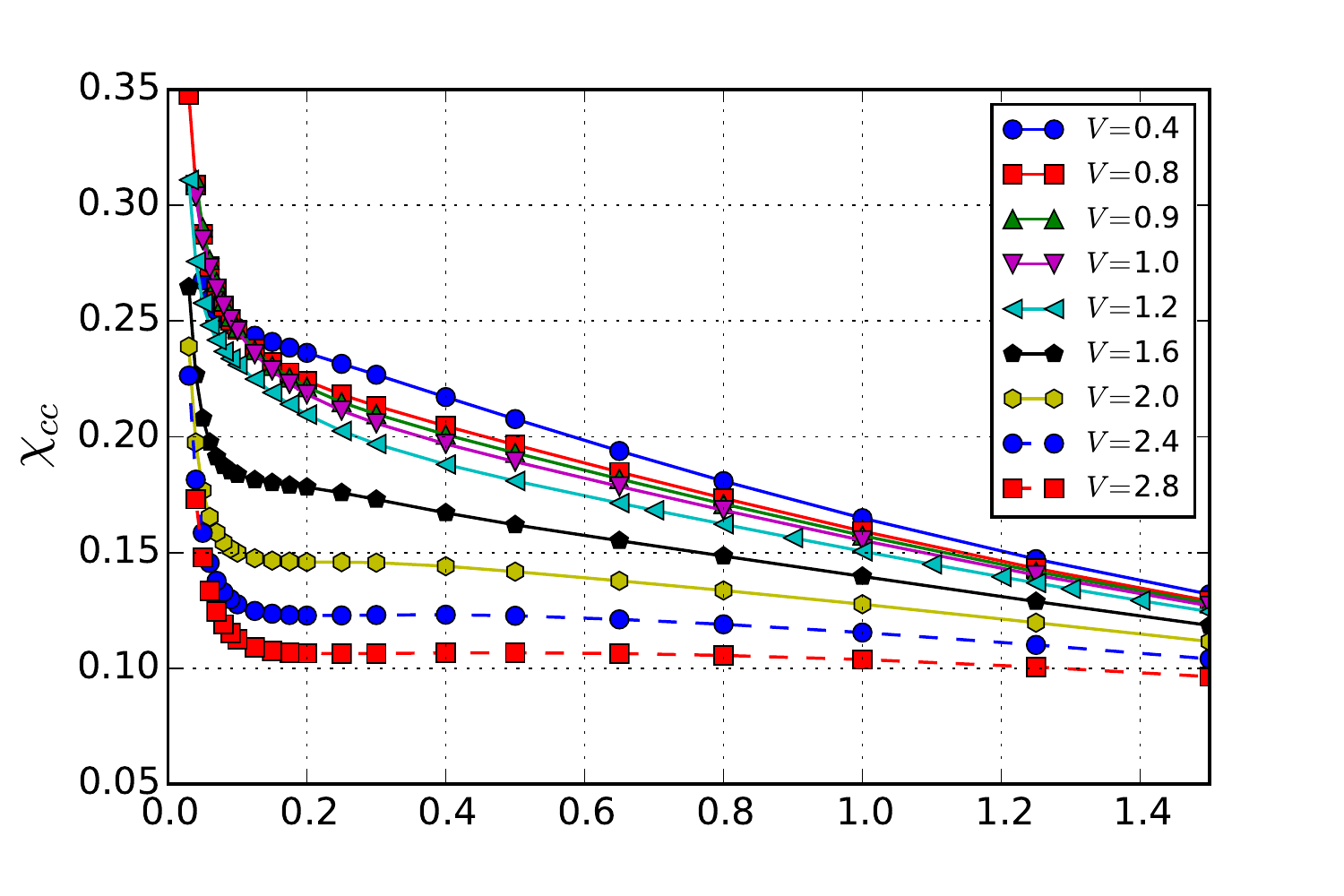,height=5.5cm,width=8.0cm,angle=0,clip} \\
\psfig{figure=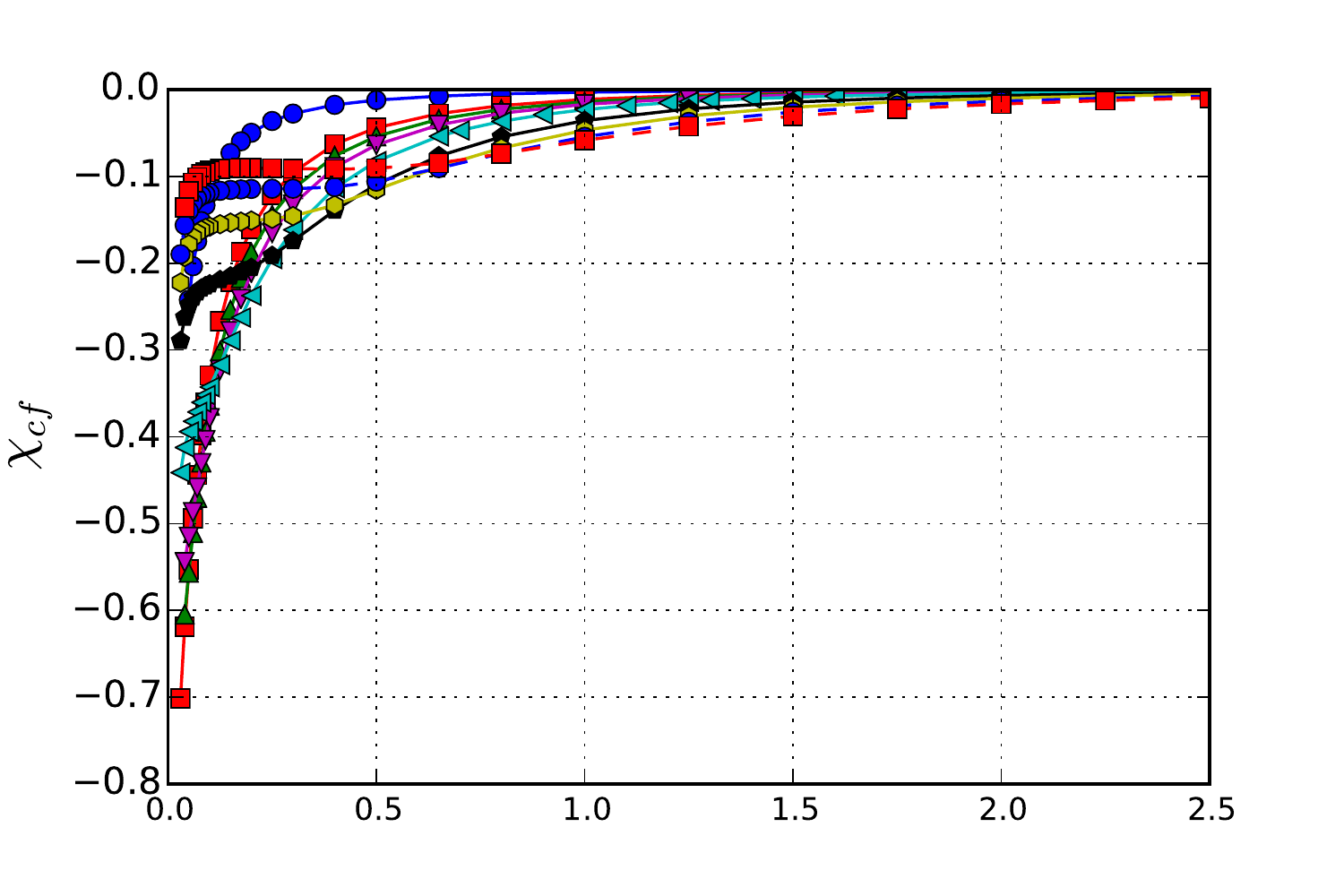,height=5.5cm,width=8.0cm,angle=0,clip=16.0cm} \\
\psfig{figure=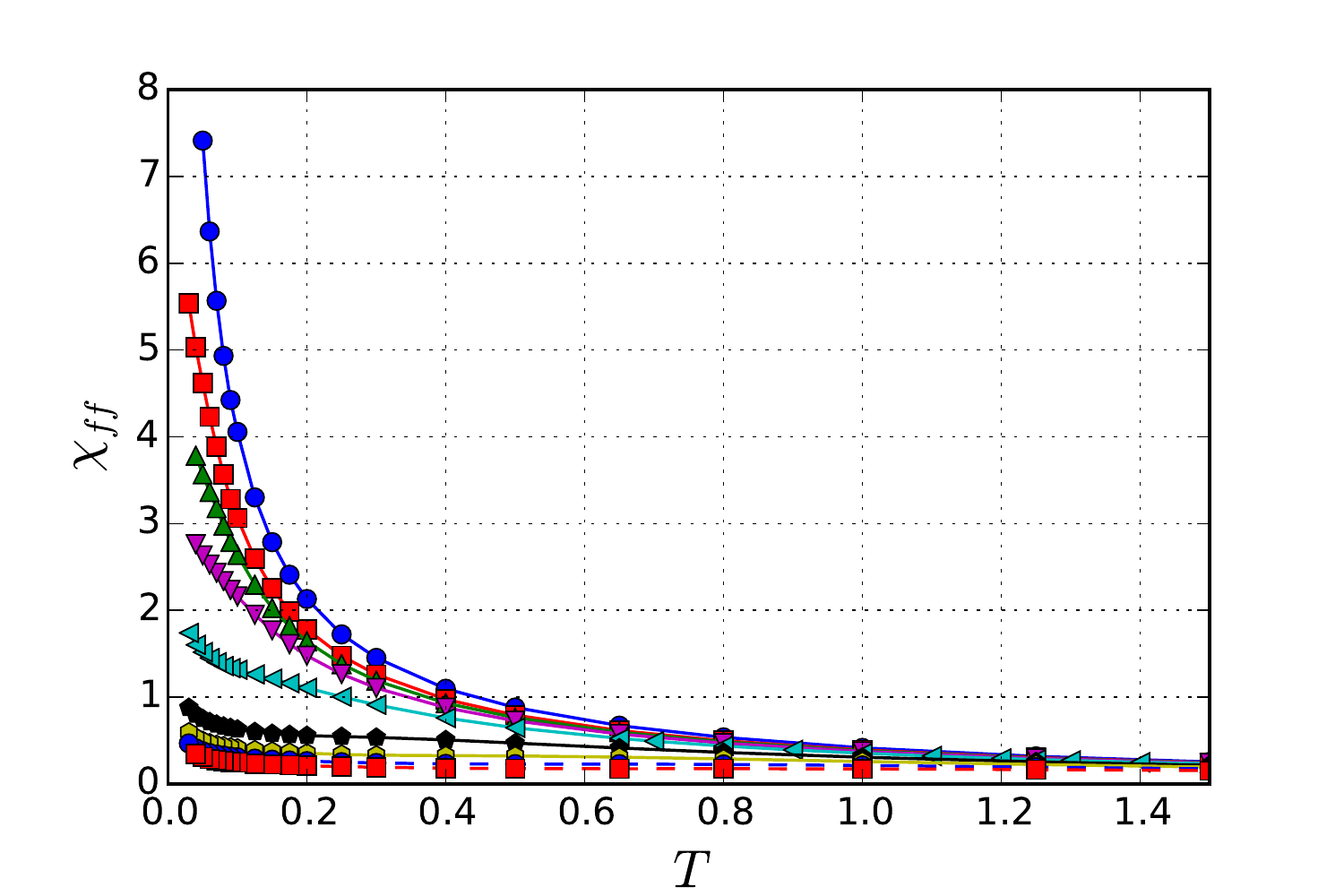,height=5.5cm,width=8.0cm,angle=0,clip} \\
\caption{(Color online) Temperature and hybridization ($V$) evolution of the three components of the local magnetic susceptibility for $\langle n_{f} \rangle=0.9$. Weak hybridization induces decoupling between the conduction and local $f$-electrons, which exhibit Pauli and Curie behaviors, respectively. Strong hybridization leads to the collective Kondo screening manifested in the gradual flatness of $T$-dependence of $\chi_{ff}$. }
\label{chis}
\end{figure}

To gain some insight into the universal Knight shift anomaly reported in experiment, a previous study has employed Determinant Quantum Monte Carlo (DQMC)~\cite{blankenbecler81} and shown that the different temperature evolution of orbital-dependent magnetic susceptibilities plays a key role. In particular, they were able to derive the universal behavior of the Knight shift anomaly below a crossover temperature $T^{\ast}$, in qualitative agreement with experiments~\cite{MJ2014}. However, a shortcoming of this previous study is that it only applies to the half-filled PAM due to the infamous fermionic sign problem. Here we extend previous investigations to the doped system by means of dynamical cluster approximation (DCA)~\cite{DCAreview}. As an extension of the dynamical mean-field theory (DMFT)~\cite{DMFTreview}, DCA represents the bulk lattice problem by a finite number of cluster degrees of freedom embedded in a self-consistent mean-field host via a coarse-graining procedure of the Green's function, in which the Brillouin zone is divided into $N_{c}$ patches and the self-energy is assumed to be constant on these patches. In this way, DCA deals with all correlations within the cluster, while the longer range correlations outside the cluster are described in the mean-field level.  This gives an approximation of the thermodynamic limit and the exact solution of the lattice model can be reproduced in the limit of infinite cluster size. In practice, the most time-consuming part of the DCA self-consistent loop is the cluster solver, which includes either perturbative techniques such as fluctuation-exchange approximation or nonperturbative techniques such as quantum Monte Carlo (QMC), exact diagonalization etc. One widely used approach is the continuous-time QMC (CT-QMC), which is based on a diagrammatic expansion of the partition function to all orders~\cite{CTQMCreview}. Throughout this paper, owing to its accuracy and efficiency, we will adopt a particular version of CT-QMC, namely the continuous-time auxiliary-field (CT-AUX) algorithm, which is based on an interaction expansion combined with an auxiliary-field decomposition of the interaction vertices~\cite{CTAUX}.

In the two-fluid theory~\cite{twofluid2004,twofluid2008,twofluid2008a,twofluid2012,twofluid2016}, the nuclear moment $\vec I$ interacts with both the conduction and localized $f$-electron spins, $\vec S_{i}^c$ and $\vec S_{i}^f$, via hyperfine interactions, ${\cal H}_{\rm hyp}=  \vec{I_i}\cdot (A\vec S_{i}^c+B\vec S_{i}^f)$, where $A$ and $B$ are their respective hyperfine coupling tensors and $\vec \sigma$ are the Pauli matrices. If the electronic spins are polarized via an external magnetic field ${\rm H}$, then $S_{i}^c=(\chi_{cc}+\chi_{cf}){\rm H}$ and $S_{i}^f=(\chi_{cf}+\chi_{ff}){\rm H}$, where $\chi_{cc}, \chi_{cf}, \chi_{ff}$ denote three components of the susceptibilities, so that the total susceptibility and Knight shift are
\begin{eqnarray}
\chi&=&\chi_{cc}+2\chi_{cf}+\chi_{ff}, \nonumber \\ K &=&
A\chi_{cc}+(A+B)\chi_{cf}+B\chi_{ff}+K_{0}, \label{eq:Kchi}
\end{eqnarray}
where $K_{0}$ is a temperature independent term arising from orbital and diamagnetic contributions to $K$. The conventional normal Fermi liquid behavior $K\sim \chi$ is reproduced if $A=B$. As discussed in previous work~\cite{MJ2014}, if $A\neq B$, the different weights and temperature dependency of the three components will result in an anomalous deviation from the linear relation between $K$ and $\chi$ below a temperature scale $T^{\ast}$, below which $\chi_{cf}$ becomes significant and $\chi_{cc}$ is no longer constant. This marks the onset of the effective hybridization between the conduction and $f$-electrons and gives rise to the emergence of the Knight shift anomaly.

Before proceeding, we remark that we focus on the particular parameters: $U/t=4$, $N_{c}=16$, $\langle n_{c} \rangle=1$. The hyperfine couplings are chosen as $A=0.2$ and $B=1.0$~\cite{ABchoice}. Because the Knight shift anomaly usually occurs at relatively high temperatures, it is reasonable that the correlation strength $U$ does not play a significant role. Besides, the local susceptibilities presented do not differ much from their uniform ($\mathbf{q}=0$) counterparts, which implies that the DCA cluster size plays a minor role as well. We have confirmed that other $U$, $N_{c}$, and even different band structures e.g., finite next nearest-neighbor hoppings $t'$ for conduction electrons and/or hybridization between orbitals, do not modify the results throughout the paper qualitatively. The fixed $\langle n_{c} \rangle=1$ is aimed to explicitly control the $f$-orbital occupancy, which is difficult to be realized for systems with fixed total occupancy. In fact, we found that the essential physics is not altered in spite of the charge redistribution between conduction and $f$ electrons.

In Fig.~\ref{chis}, we follow the method in~\cite{MJ2014} and calculate the temperature evolution of the three components, $\chi_{cc}, \chi_{cf}, $ and $\chi_{ff}$, for varying hybridization parameter $V$. For weak hybridization (small $V$), the conduction electrons only weakly interact with the $f$-electron moments. Therefore, at low temperatures, $\chi_{cc}$ exhibits weakly $T$-dependent Pauli-like behavior~\cite{footnote2}, while $\chi_{ff}$ shows a Curie-like divergence, as qualitatively captured in Figs.~\ref{chis}(a) and \ref{chis}(c). For strong hybridization (large $V$), the conduction electron spins and the $f$-electron moments tend to anti-align and form spin singlets, as manifested in the gradual flatness of the temperature evolution of $\chi_{ff}$. For $U/t=4$ it is known \cite{PAMRTS} that the transition from antiferromagnetism to Kondo singlets occurs at $V/t \gtrsim 1.2$. DCA tends to overestimate the transition point due to the presence of the mean-field host for finite cluster size, but is expected to recover the correct results in the limit of infinite cluster size. Fig.~\ref{chis}(b)  shows the behavior of he inter-orbital susceptibility $\chi_{cf}$, which is negative due to the anti-alignment between the conduction and $f$-electron spins. As expected, a stronger hybridization leads to a larger $|\chi_{cf}|$.

\begin{figure}
\psfig{figure=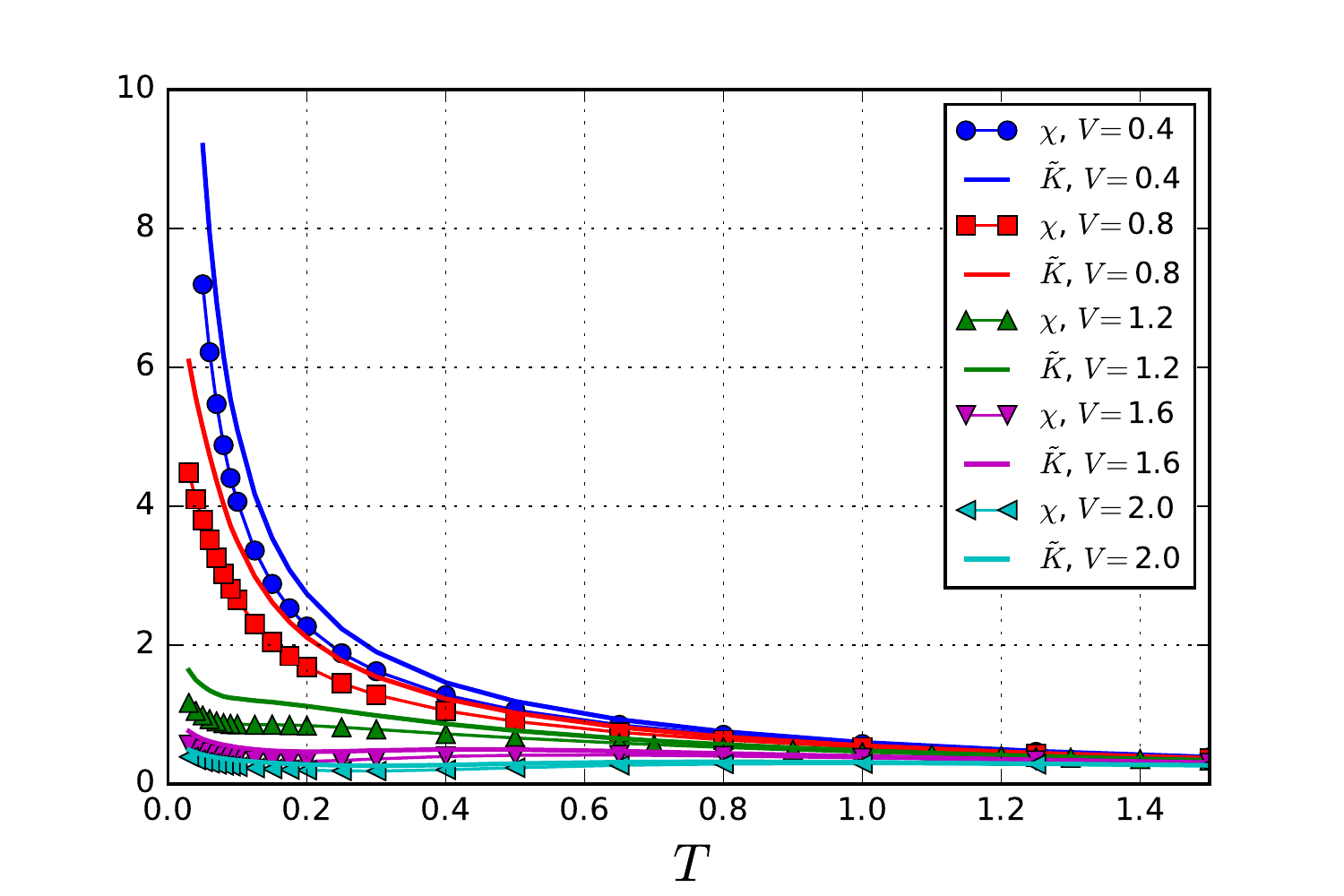,height=5.5cm,width=8.0cm,angle=0,clip} \\
\caption{(Color online) Comparison of the temperature dependence of the renormalized Knight shift $\tilde K = (K-K_{0 \,{\rm eff}})/B_{\rm eff}$ and total susceptibility $\chi$ for $\langle n_{f} \rangle=0.9$ shows their deviation below a certain temperature. }
\label{Kchi_compare}
\end{figure}

\begin{figure}
\psfig{figure=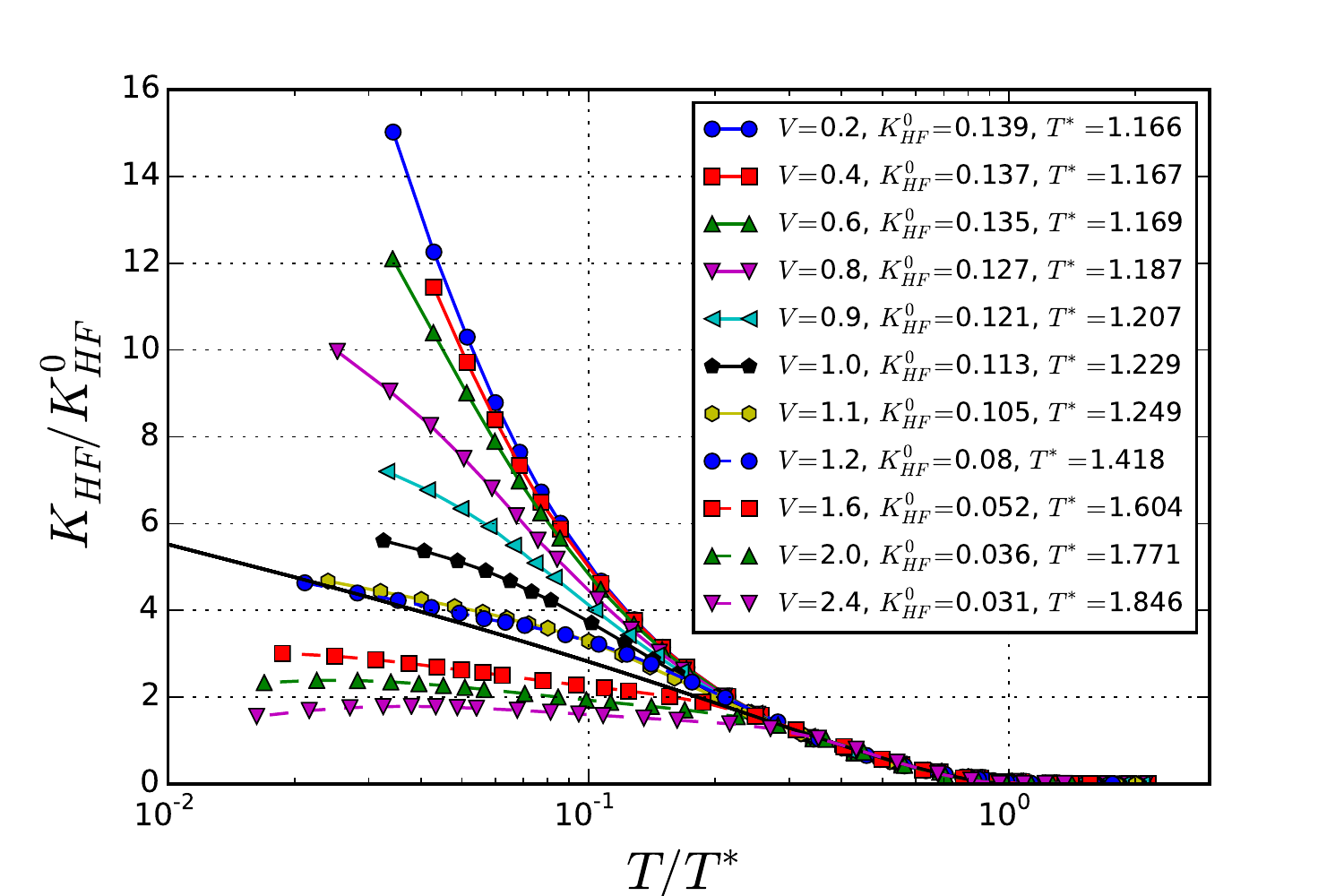,height=5.5cm,width=8.0cm,angle=0,clip} \\
\caption{(Color online) Universal logarithmic scaling of $K_{\rm HF}$ with decreasing temperature below a hybridization dependent  temperature $T^{\ast}$ for a wide range of hybridization for $\langle n_{f} \rangle=0.9$. The low temperature behavior of $K_{\rm HF}$ after the breakdown of the scaling below $\sim 0.25 T^\ast$ can be classified into three different categories, reflecting different situations of the competition between the localized and itinerant behaviors of the $f$-electrons.
}
\label{K_HF_scaling}
\end{figure}

Using the same procedure in previous experimental analysis~\cite{KentCurro} and DQMC studies~\cite{MJ2014}, we calculate the Knight shift $K$ and the total susceptibility $\chi$ in Eq. (2) using particular choices~\cite{ABchoice} of $A$ and $B$ and fit them with a straight line, $K=B_{\rm eff} \, \chi + K_{0\, {\rm eff}}$. Fig.~\ref{Kchi_compare} compares $\chi$ and $\tilde K = (K-K_{0 \,{\rm eff}})/B_{\rm eff}$ as a function of temperature. The deviation from the linear relation is evident at low temperatures for a wide range of hybridization, signifying the emergence of the Knight shift anomaly. The two-fluid model argues that the Knight shift anomaly, $K_{\rm HF}=K-(K_{0\,{\rm eff}} + B_{\rm eff} \chi)$, is proportional to the magnetic susceptibility of the heavy-electron fluid, and exhibits a {\it universal} logarithmic divergence with decreasing temperature below $T^{\ast}$~\cite{twofluid2008}:
\begin{equation}\label{universal1}
  K_{\rm HF}(T)=K_{\rm HF}^{0}(1-T/T^{\ast})^{3/2}[1+\ln(T^{\ast}/T)],
\end{equation}
where $K_{\rm HF}^{0}$ and $T^{\ast}$ are material-dependent constants. Figure~\ref{K_HF_scaling} shows the fit to the calculated $K_{\rm HF}$ using the above scaling formula for a wide range of hybridization. The fitting parameters are $K_{\rm HF}^0$ and $T^\ast$. The excellent agreement (for $0.25 T^*<T<T^*$) demonstrates that the Knight shift anomaly indeed displays a universal logarithmic scaling below a $V$-dependent temperature $T^{\ast}$. This provides a microscopic demonstration of the predicted scaling.

\begin{figure}
\psfig{figure=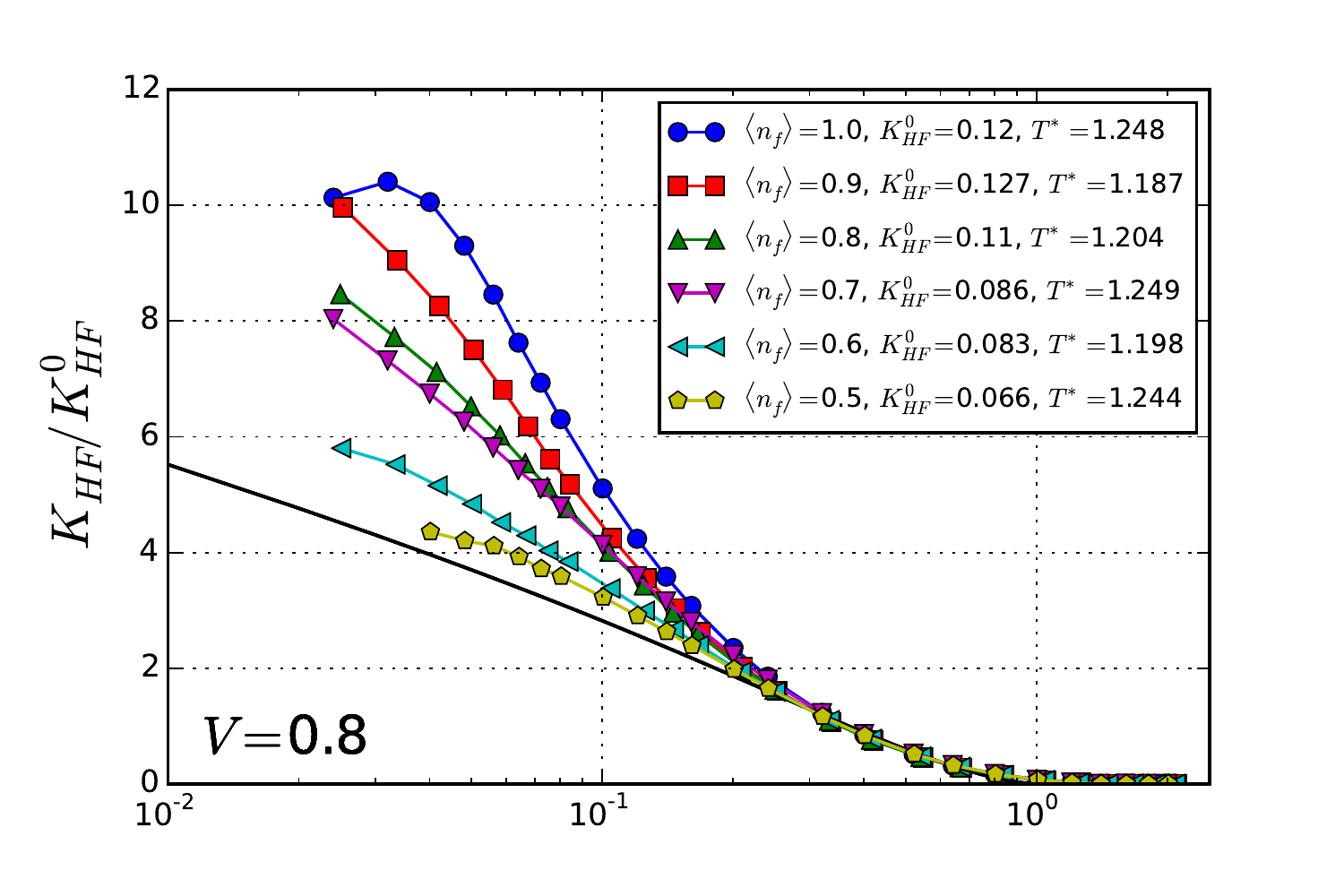,height=5.0cm,width=8.0cm,angle=0,clip} \\
\psfig{figure=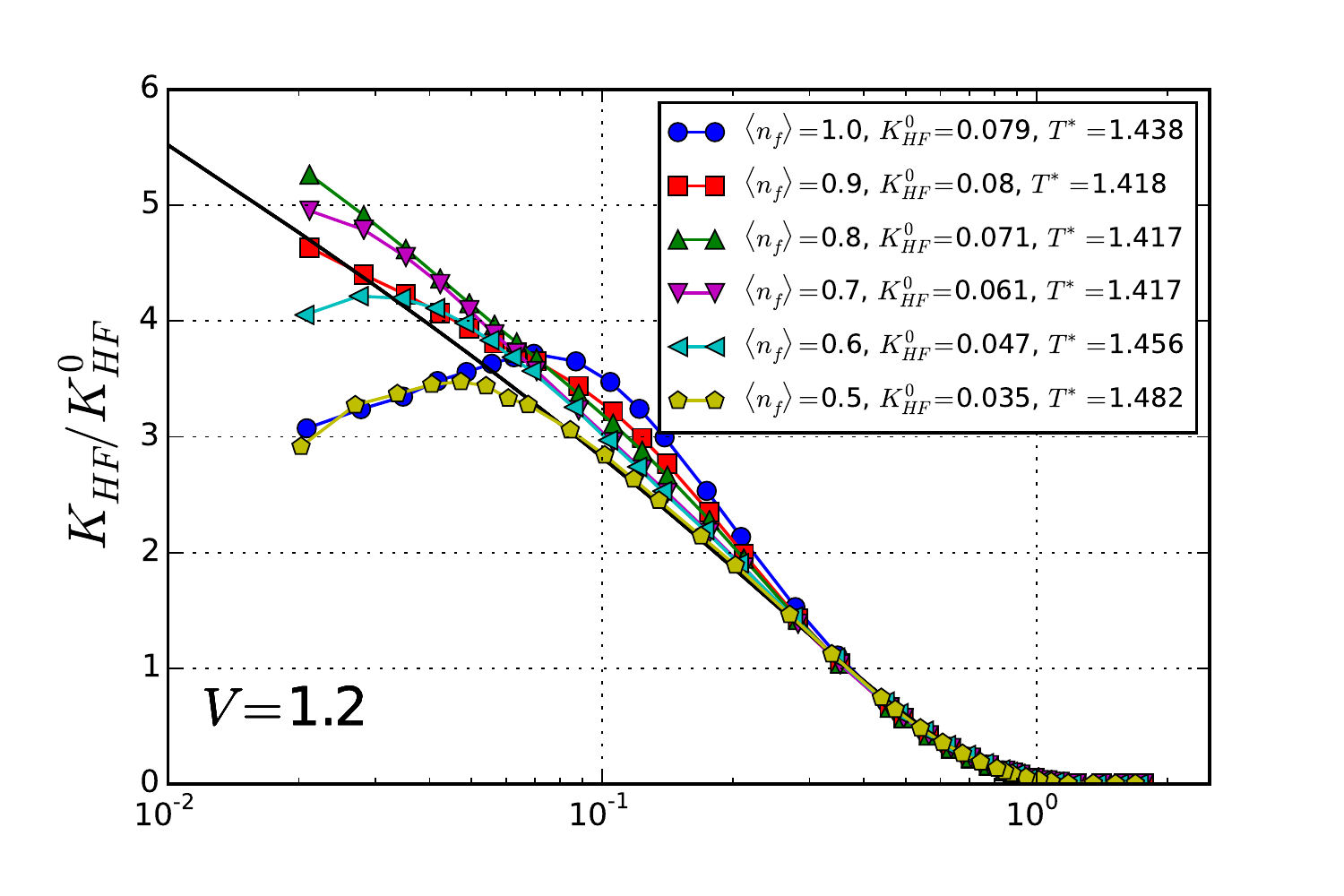,height=5.0cm,width=8.0cm,angle=0,clip=16.0cm} \\
\psfig{figure=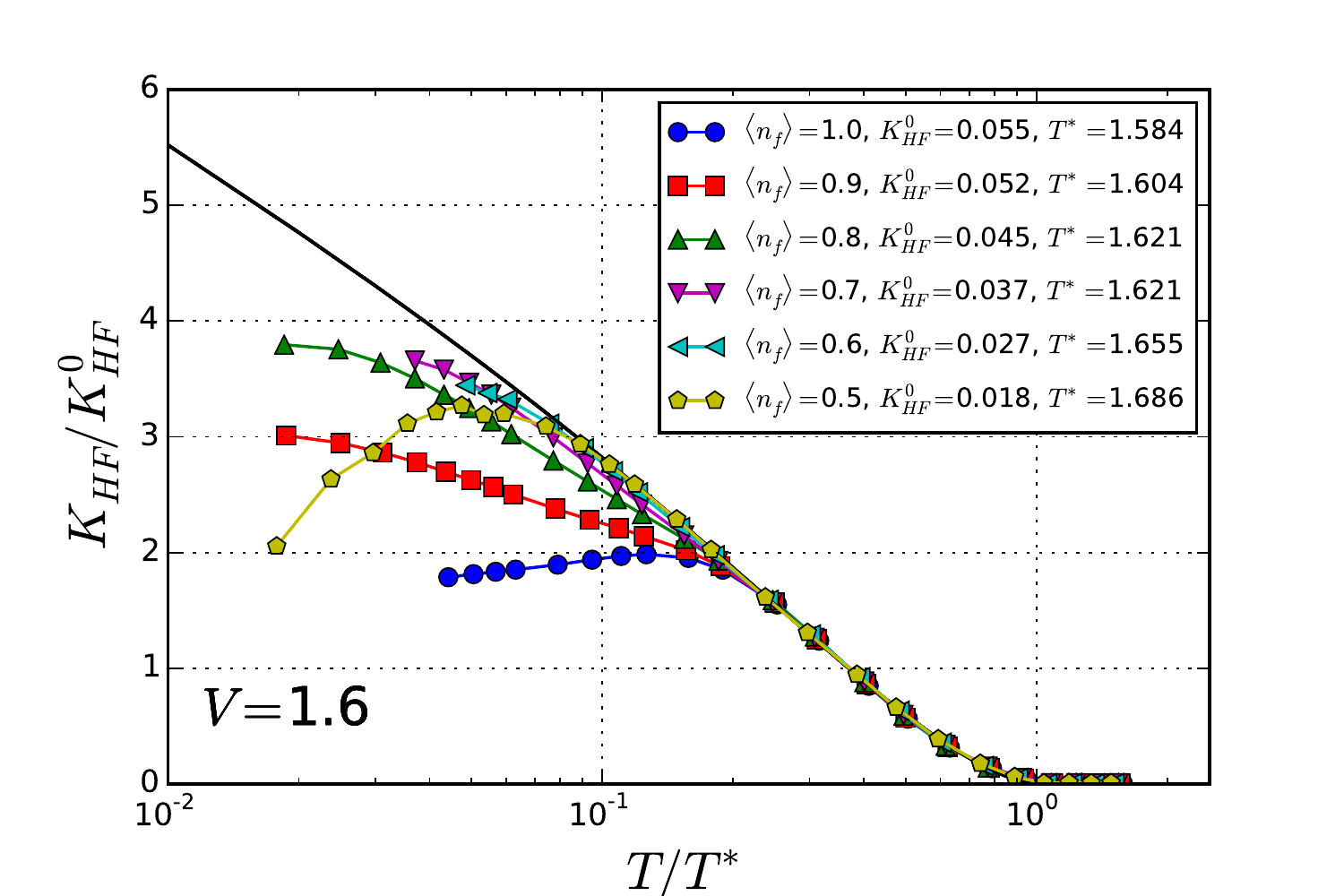,height=5.0cm,width=8.0cm,angle=0,clip} \\
\caption{(Color online) Universal logarithmic divergence of $K_{\rm HF}$ as a function of temperatue for different $f$-electron occupancy $\langle n_{f} \rangle$. $K_{\rm HF}$ in three regimes of low temperature behaviors similar to Fig.~\ref{K_HF_scaling}. Doping strongly modifies the behavior of $K_{\rm HF}$ in comparison with the scaling curve. A crossover is seen at $V=1.2$, possibly corresponding to the transition between heavy-fermion and strong mixed-valence regimes. }
\label{universal_dope}
\end{figure}

Our results in Fig.~\ref{K_HF_scaling} reveal several noticeable features (away from half filling). Evidently, the scaling in the Knight shift anomaly, $K_{\rm HF}$, only applies between $T^*$ and a lower temperature scale $T_{0} \sim 0.25 T^\ast$. This is consistent with experimental observations and reflects the intervene of other effects at temperatures below $T_0$ \cite{twofluid2012}. The low temperature behaviors of $K_{\rm HF}$ differ distinctively between weak and strong hybridizations. In particular, at small $V$ and absent antiferromagnetism, $K_{\rm HF}$ keeps increasing and the deviation from the scaling formula originates in part from the interplay between the residual $f$-moments and the conduction electrons. In the presence of long-range antiferromagnetic order, the Knight shift anomaly is typically suppressed, owing to the competition between the heavy electron formation and the localization caused by the magnetic ordering. This is termed as relocalization of heavy electrons, as has been observed in CeRhIn$_5$, CePt$_2$In$_7$ and other heavy-fermion antiferromagnet \cite{relocalization,KentCurro}. In either case, the $f$-electron moments remain partially screened and partially localized and one expects a continuing competition between the itinerant and localized behavior, causing possible coexistence of long-rang antiferromagnetism (or a spin liquid in the absence of long-range order) and unconventional superconductivity \cite{twofluid2012}. While at large $V$, $K_{\rm HF}$ is seen to saturate below $T_0$. The constant behavior at low temperatures reflects complete hybridization of the $f$-electron moments and the ground state is a heavy Fermi liquid. Between these two regimes, one finds a minimal deviation of $K_{\rm HF}$ from the two-fluid scaling at an intermediate $V/t \sim 1.2$. This corresponds roughly to the quantum critical point between antiferromagnetism and the Fermi liquid, suggesting that the two-fluid scaling is less intervened by low temperature ``orders''. We note that the onset temperature $T^\ast$ increases with the hybridization $V$, reflecting the enhancement of the $f$-electron coherence. On the other hand, the prefactor, $K_{\rm HF}^0$, decreases with the hybridization. In the two-fluid model, $K_{\rm HF}^0$ is inversely proportional to $T^*$ \cite{twofluid2012}, which may partially explain the change in $K_{\rm HF}^0$.

\begin{figure}
\psfig{figure=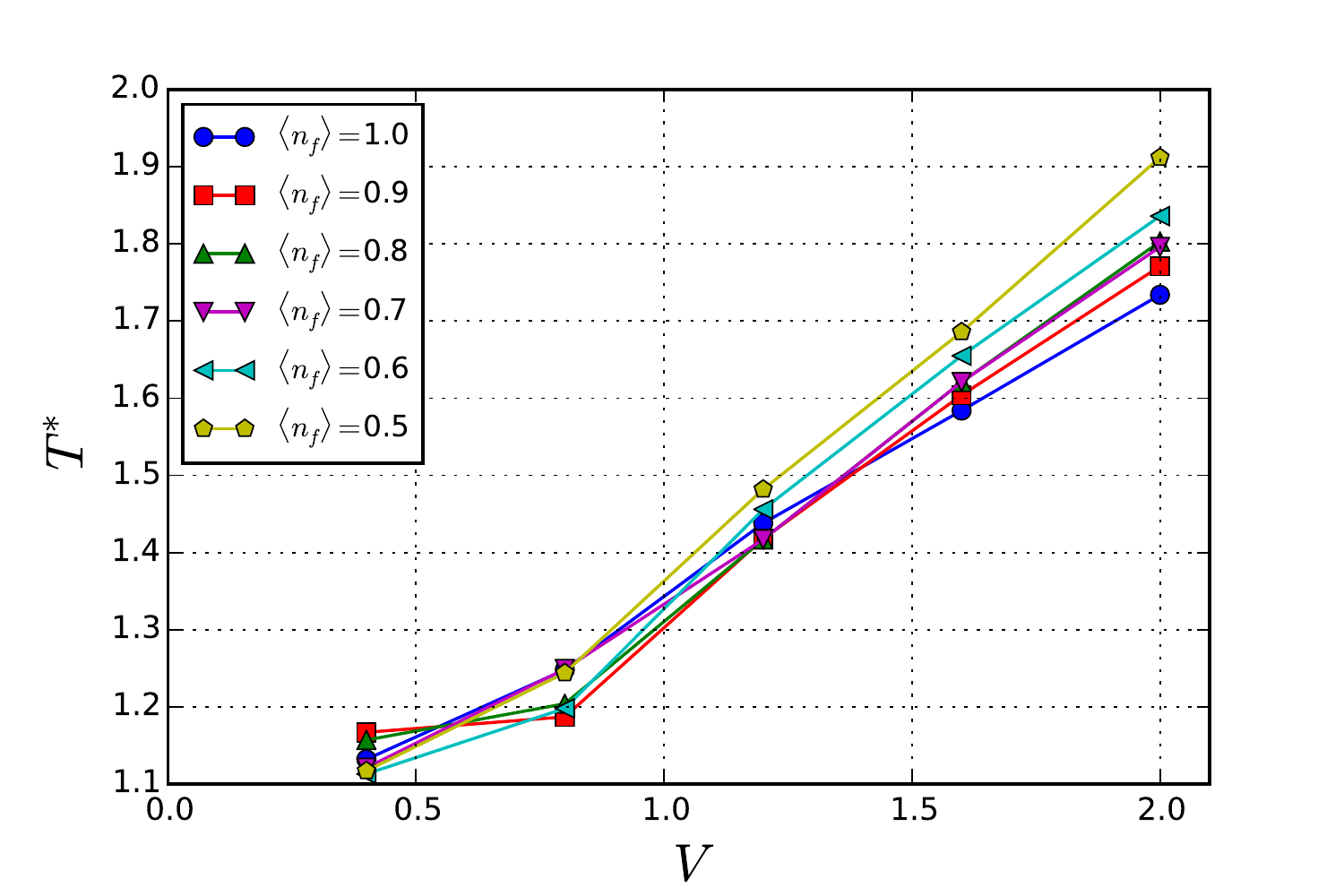,height=5.5cm,width=8.0cm,angle=0,clip} \\
\caption{(Color online) Evolution of $T^{*}$ with doping and hybridization based on the results in Fig.~\ref{universal_dope}. For all doping levels including half-filling, $T^{\ast}$ grows rapidly with $V$ but only weakly with the doping.}
\label{Ts}
\end{figure}

Fig.~\ref{universal_dope} provides a systematic study of the Knight shift anomaly for varying $\langle n_{f} \rangle$ in all three regimes of hybridizations. Similar to Fig.~\ref{K_HF_scaling}, $K_{\rm HF}$ displays universal scaling behavior below $T^*$ down to a breakdown temperature $T_{0}$. For weak hybridization, $K_{\rm HF}$ approaches gradually to the scaling formula at low temperatures with decreasing $\langle n_{f} \rangle$, indicating the weakening of the local moment effect away from half filling.
For intermediate and strong hybridizations, $K_{\rm HF}$ below $T_0$ first approaches the universal scaling but then deviates again. This nonmonotonic variation with $\langle n_{f} \rangle$ reflects the crossover from the heavy-fermion regime ($\langle n_{f} \rangle=1$) to the mixed-valence regime. In particular, it suggests two different Fermi liquid states for $\langle n_{f} \rangle=1$ and $\langle n_{f} \rangle \ll 1$ at strong hybridization, probably separated by an intermediate non-Fermi liquid phase at $\langle n_{f} \rangle \sim 0.75$. Interestingly, for all doping levels including half filling, the onset temperature of the Knight shift anomaly, $T^{\ast}$, grows rapidly with the hybridization parameter $V$ as expected, but only weakly on $\langle n_{f} \rangle$ as shown in Fig.~\ref{Ts}. This suggests that the variation of $\langle n_{f} \rangle$ plays a less important role in determining the magnitude of the hybridization compared to $V$.

\begin{figure}
\psfig{figure=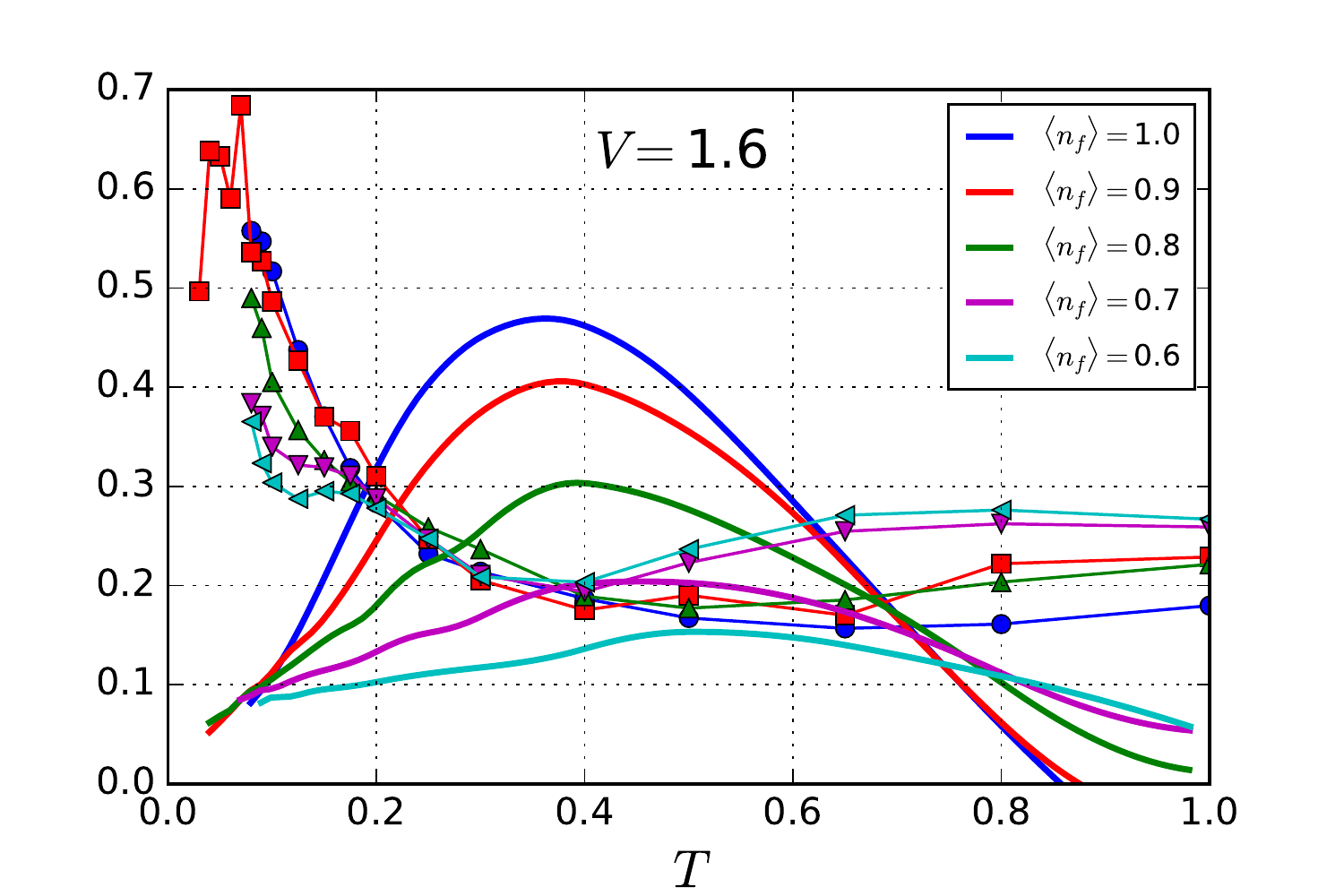,height=5.5cm,width=8.0cm,angle=0,clip} \\
\caption{(Color online) Temperature derivative of the quasiparticle scattering rate $-d$(Im$\Sigma_{f}$)/$d\ln T(i\omega=\pi T)$ and the quasiparticle peak height for $f$-electrons as a function of temperature, showing a common temperature scale at $T^{\ast}/4$, much lower than the onset temperature of the Knight shift anomaly.}
\label{compare_dos_vs_Imsigma}
\end{figure}

To better understand the physics of $T^*$, we calculate the density of states (DOS) and the self-energy to examine their connection with the appearance of the Knight shift anomaly. In previous two-fluid analysis of existing experimental data on CeCoIn$_5$ and many other heavy-fermion compounds~\cite{twofluid2008a}, it has been suggested that $T^*$ provides a unified temperature/energy scale for the magnetic, transport and spectral properties, which should be reflected in the Knight shift, resistivity and DOS, respectively. The temperature dependence of the resistivity can be qualitatively captured by the temperature derivative of the quasiparticle scattering rate, Im$\Sigma_{f}(\omega=0)$. As pointed out in~\cite{Shim2007}, the peak position of $-d$(Im$\Sigma_{f}$)/$d\ln T(\omega=0)$ gives a good estimate of the coherence temperature in the resistivity. On the other hand, previous calculations of CeIrIn$_5$~\cite{Shim2007,twofluid2008} have also shown that the universal scaling can be manifested in the temperature scaling of the quasiparticle peak. Fig.~\ref{compare_dos_vs_Imsigma} compares $-d$(Im$\Sigma_{f}$)/$d\ln T(i\omega=\pi T)$ and the quasiparticle DOS of $f$-electrons as a function a temperature for different $\langle n_{f} \rangle$. We see clearly that the peak of quasiparticle scattering rate roughly agrees with the temperature below which the quasiparticle peak shows a rapid increase. However, these occur at about $T^*/4$, much lower than the onset of the Knight shift anomaly, which is in distinct difference from the experimental observation in CeCoIn$_5$ and many other compounds. However, we should note that there does exist few example such as CeCu$_2$Si$_2$ and UBe$_{13}$, where the resistivity peak appears at a much lower temperature than $T^*$ estimated from the NMR Knight shift~\cite{twofluid2008a}. There are two possible reasons for this discrepancy: either accidental due to the involvement of some other effects, or representing a particular kind of heavy-fermion physics different from that of CeCoIn$_5$. In either case, it would be important to clarify the conditions for which the three quantities exhibit similar/different $T^*$. We leave this for future investigations.

To summarize, we extend previous theoretical study of the NMR Knight shift in the half-filled periodic Anderson model to the doped case using the DCA method. Our simulations show that the universal scaling of the Knight shift anomaly persists in the whole doping and hybridization regime and represents a robust property of the Anderson lattice. Our work provides a plausible basis for developing a microscopic understanding of the phenomenological two-fluid model. However, it also indicates that some essential physics is missing in the current model calculations, as it cannot reproduce the common $T^*$ for the magnetic, transport and spectral properties observed in CeCoIn$_5$ and many other heavy-fermion compounds. This reveals a missing piece in explaining the two-fluid phenomenology in the current model calculations, which may be the key for a thorough solution of the heavy-fermion problem.

\begin{acknowledgments}
We acknowledge the Swiss National Supercomputing Center (CSCS) for computing facility support. Y.Y. was supported by the National Natural Science Foundation of China (Grant No. 11522435) and the Youth Innovation Promotion Association CAS.
\end{acknowledgments}


\bibliography{CurroBibliography}

\begin{thebibliography}{40}
\bibitem{ABchoice}
We adopt $B=1$ so that only $A/B$ is varied. It is found that the specific values of hyperfine interactions $A$ and $B$ do not modify any results qualitatively.
\bibitem{footnote2}
The rapid increase of $\chi_{cc}$ at low temperature is partly due to the half-filled conduction band in that the van-Hove singularity in the density of states of a tight binding model such as PAM with nearest-neighbor hopping in square lattices induces the lack of flatness of $\chi_{cc}(T\rightarrow 0)$ expected from Pauli susceptibility behavior.
\end{thebibliography}

\end{document}